\input phyzzx
\voffset = -0.4in 
\rightline {QMW-TH-96-21.}
\title {SYMPLECTIC EMBEDDINGS, THE PREPOTENTIALS AND AUTOMORPHIC FUNCTIONS OF 
$\textstyle SU(1, n)\over \textstyle U(1)\otimes SU(n)$}
\author{W. A. Sabra$^{a}$,\foot{e-mail: uhap012@vax.rhbnc.ac.uk}
S. Thomas$^{b}$\foot{e-mail: s.thomas@qmw.ac.uk} and N. Vanegas$^{b}$
\foot{e-mail: n.vanegas@qmw.ac.uk}}
\address{$^{a}$ Physics Department,
Royal Holloway and Bedford New College,\break
University of London, Egham Surrey, \break
TW20 OEX, United Kingdom.}
\address{$^{b}$ Physics Department,\break
Queen Mary and Westfield College,\break
Mile End Road, London E1 4NS,\break
United Kingdom.}
\abstract{The holomorphic homogeneous prepotential 
encoding the special geometry of the special K\"ahler manifolds
${\textstyle SU(1,n)\over \textstyle U(1)\otimes SU(n)}$
is constructed using the symplectic embedding of the isometry
group $SU(1,n)$ into $Sp(2n+2,{\hbox{\bf R}})$.
Also the automorphic functions of these manifolds are discussed.}
\endpage
\REF\trieste{B. de Wit and A. Van Proeyen, {\it Special geometry and
symplectic transformation}, lectures delivered by A. Van Proeyen at the
String workshop on String theory. Trieste, April 1994, hep-th/9510186.}
\REF\one{B. de Wit, P. G. Lauwers, R. Philippe, Su S. Q and A. Van 
Proeyen, {\it Phys. Lett.} {\bf B134} (1984) 37; 
B. de Wit and A. Van Proeyen, {\it Nucl. Phys.}
{\bf B245} (1984) 89.}
\REF\two{E. Cremmer, C. Kounnas, A. Van Proeyen, J. P. Derendinger,
S. Ferrara, B. de Wit and L. Girardello, {\it Nucl. Phys.} 
{\bf B250} (1985) 385.}
\REF\three{B. de Wit, P. G. Lauwers and A. Van Proeyen, 
{\it Nucl. Phys}. {\bf B255} (1985) 569.}
\REF\four{E. Cremmer and A. Van Proeyen, {\it Class. Quantum Grav.}
{\bf 2} (1985) 445.}
\REF\five{S. Ferrara and A. Van Proeyen, {\it Class. Quantum Grav.}
{\bf 6} (1989) 124.}
\REF\six{A. Strominger, {\it Commun. Math. Phys.} {\bf 133}
(1990) 163.}
\REF\seven{L. Castellani, R. D'Auria and S. Ferrara, {\it Phys. Lett.}
{\bf B241} (1990) 57.}
\REF\eight{R. D'Auria, S. Ferrara and P. Fr\`e, {\it Nucl. Phys.} 
{\bf B359} (1991) 705.}
\REF\nine{S. Cecotti, 
S. Ferrara, and L. Girardello, {\it Int. J. Mod. Phys.}
{\bf 4} (1989) 2475; {\it Phys. Lett.} {\bf  B213} (1988) 443.}
\REF\ten{S. Ferrara and A. Strominger, in Strings
'89, ed. R. Arnowitt, R. Bryan, M. J. Duff, D. Nanopulos and C. N. Pope,
World Scientific, Singapore, (1990) 245.} 
\REF\eleven{L. Dixon, V. Kaplunovsky and J. Louis, {\it Nucl. Phys.}
{\bf B329} (1990) 27.}
\REF\twelve{P. Candelas and X. de la Ossa, 
{\it Nucl. Phys.} {\bf B355} (1991) 455; P. Candelas,  X. de la Ossa, 
P. Green and L. Parkers, 
{\it Nucl. Phys.} {\bf B359} (1991) 21.}
\REF\fre{P. Fr\`e, P. Soriani, {\it Nucl
Phys} {\bf B371} (1992) 659.}
\REF\GZ {M. K. Gaillard and B. Zumino, {\it Nucl. Phys.} {\bf B193}
(1981) 221.}
\REF\eighteen{S. Ferrara, C. Kounnas,  D. L\"ust and F. Zwirner,
{\it Nucl. Phys.} {\bf B365} (1991) 431.}
\REF\lust{G. L. Cardoso, D. L\"ust and T. Mohaupt, {\it  Nucl. Phys.}
{\bf B450} (1995) 115.}
\REF\van{A. Ceresole, R. D'Auria, S. Ferrara and A. Van Proeyen, 
{\it Nucl. Phys.} {\bf B444} (1995) 92.}
\REF\wit{B. de Wit, V. Kaplunovsky, J. Louis and D. L\"ust, {\it  Nucl.
Phys.}{\bf B451} 
(1995) 53; I. Antoniadis, S. Ferrara, G. Gava, 
K. S. Narain and T. R. Taylor, {\it Nucl. Phys.} {\bf B447} (1995) 35.}
\REF\br{S. Ferrara, L. Girardello and M. Porrati, hep-th/9510074.} 
Special geometry has emerged as an important structure in the study
of supergravity and superstring theories [\trieste-\twelve]. 
This geometric structure arises in the study 
of the moduli spaces of $N=1$ compactifications on Calabi-Yau manifolds, 
$N=2$ supergravity theory and appears as a consequence of 
$(2,2)$ worldsheet superconformal field theory or $N=2$ supersymmetry 
in the target space. 

The concept of special K\"ahler geometry has first 
appeared in the physics literature in the study of the coupling 
of an $n$ vector multiplets to four dimensional $N=2$ supergravity
[\one,\two,\three].  
The lagrangian for the coupling of the vector multiplets was derived using the

superconformal tensor calculus [\one]. In addition to the hypermultiplets, the
theory is 
described by 
introducing $(n+1)$ vector multiplets with scalar components $X^I,$ 
$(I=0,\cdots, n)$, where the extra multiplet labelled by 0 contains  
the graviphoton. The couplings were found to be described 
in terms of a holomorphic 
function $F$, referred to as the prepotential, of degree two in terms of the 
scalar fields defining an $n$-dimensional complex hypersurface 
defined by the gauge fixing condition
$$i(X^I\bar F_I -F_I\bar X^I)=1.\eqn\co$$ 
Letting the $X^I$ be proportional to holomorphic sections $Z^I(z)$ of a 
projective $(n+1)$-dimensional space, where $z$ is a set of $n$ 
complex coordinates, then the $z$ coordinates parametrize a K\"ahler space
with metric $g_{\alpha\bar\beta}=\partial_\alpha\partial_{\bar\beta}
K$, where $K$, the K\"ahler potential is expressed by 
$$\eqalign{&K=-\log i(Z^I\bar F_I(\bar Z) -F_I(Z)\bar Z^I)\cr
&=-log\Big(i<\Omega\vert\bar\Omega>\Big)=-\log i\pmatrix{Z^I&F_I(Z)}
\pmatrix{{\bf 0}&{\bf 1}\cr -{\bf 1}&{\bf 0}}\pmatrix {{\bar Z}^I\cr{\bar
F}_I(\bar Z)},\cr
& X^I=e^{K/2}Z^I.}\eqn\k$$

Special coordinates correspond to the choice
$$z^\alpha={X^\alpha\over X^0}; \qquad Z^0(z)=1,
\quad Z^\alpha(z)=z^\alpha.\eqn\madness$$

The gauge kinetic terms, expressed in terms of the matrix $\cal N$, 
corresponding to the $(n+1)$ vector 
are given by
$${\cal N}_{ij}=\bar F_{ij}+2i
{(Im F_{im})(Im F_{jn} )X^mX^n\over (Im F_{ab})X^aX^b}.\eqn\thunder$$
 
Homogeneous symmetric special manifolds  
were classified in [\four] with their corresponding holomorphic prepotentials.
In addition to four sporadic cases, the list of special K\"ahler manifolds 
includes two series, one of which is given in the form of a direct product, 
which have been proved to be the only allowed case [\five]. 
These two series are given
by
$$\eqalign{SK(n+1)=&{SU(1,1)\over U(1)}\otimes {SO(2,n)\over SO(2)
\otimes SO(n)}, \cr CP_{n-1, 1}=&{SU(1,n)\over SU(n)\otimes U(1)}.}\eqn\ser$$

The classical holomorphic prepotentials for the manifolds $SK(n+1)$ have been 
analyzed in [\fre]. There, the classical holomorphic function $F$ was 
obtained in the spirit of the method of [\GZ] in constructing 
the lagrangian of an abelian gauge theory 
with $n$ gauge fields and a set of scalar fields parametrizing a 
coset space. In this construction, the duality transformations on the 
gauge sector of the theory are parametrized by the embedding of the isometry
group 
of the coset space of the scalar fields into the group
$Sp(2n+2,{\hbox{\bf R}})$ or 
$Usp(n,n).$
Such an embedding can then be used to determine the dependence of the gauge
kinetic terms on the scalar fields.
In [\fre], the section $(X^\Lambda,F_\Lambda)$ is introduced as 
a symplectic vector. A relation between $F_\Lambda$ and $X^\Lambda$
is then determined in an appropriate embedding of the  
isometries of 
$SK(n+1)$ into $Sp(2n+2,{\hbox{\bf R}}).$

In this letter, the generalization of the results of [\fre] 
to the second infinite
series of special manifolds $CP_{n-1,1},$ the so called minimal coupling, 
is performed. We construct their holomorphic prepotentials encoding their 
special geometry and discuss their automorphic 
functions which could serve as non-perturbative superpotentials for the scalar
fields. 

The isometry group of the cosets $CP_{n-1,1}$ is given by the group 
$SU(1,n)$. As a first step in the construction of the holomorphic 
function for these manifolds, we have to construct an embedding of their 
isometry group into 
the symplectic
group $Sp(2n+2,{\hbox{\bf R}})$.

We represent an element of $SU(1, n)$ by the complex $(n+1)\times (n+1)$ 
matrix $\bf M$ 
satisfying the conditions 
$${\bf M}^\dagger\bf \eta {\bf M}=\eta, \quad \det {\bf M}=1,\eqn\mars$$ where
$\eta$ is the 
constant diagonal metric with signature $(+, -,\cdots, -)$. 
Decomposing the matrix $\bf M$ into its real and imaginary parts,
$${\bf M}={\bf U}+i{\bf V},\eqn\dec$$ then from \mars\ we obtain, 
for the real matrices ${\bf U}$ and ${\bf V},$ the following relations
$$\eqalign{{\bf U}^t\eta {\bf U}+{\bf V}^t\bf\eta {\bf V}&=\bf\eta, 
\cr{\bf U}^t\bf\eta {\bf V}-{\bf V}^t\bf\eta {\bf U}&=0.}\eqn\cybill$$
Represent an element of $Sp(2n+2,{\hbox{\bf R}})$ by the $(2n+2)\times (2n+2)$
real matrix 
$\bf\Omega$, this element satisfies
$${\bf\Omega}^t\bf L{\bf\Omega}= \bf L,\qquad \bf 
L=\pmatrix{{\bf 0}&{\bf 1}\cr -{\bf 1}&{\bf 0}},\eqn\pinhead$$
If we write $$\bf\Omega=\pmatrix{{\bf A}&{\bf B}\cr {\bf C}&{\bf D}},
\eqn\swansha$$
where the matrices $\bf A$, $\bf B,$ $\bf C$ and $\bf D$ are 
$(n+1)\times (n+1)$ matrices, then in  terms of these component matrices, 
the condition in \pinhead\ implies the following conditions 
$${\bf A}^t{\bf C}-{\bf C}^t{\bf A}={\bf 0},\quad
{\bf A}^t{\bf D}-{\bf C}^t{\bf B}={\bf I},\quad 
{\bf B}^t{\bf D}-{\bf D}^t{\bf B}={\bf 0}.\eqn\ero$$ 

An embedding of $SU(1,n)$ into the symplectic group $Sp(2n+2,{\hbox{\bf R}})$
is given 
by 
$${\bf A}={\bf U}, \quad {\bf C}=-\eta {\bf V}, \quad {\bf B}={\bf V}\eta,
\quad {\bf D}=\eta {\bf U}\eta.\eqn\com$$
Having obtained an embedding ${\bf\Omega}_e$ with matrix components 
as given in \com, the next step is to introduce the symplectic section 
$(X^\Lambda, F_\Lambda),$ where $\Lambda=0,\cdots, n$,  and demand 
that it transforms as a vector under the symplectic transformations 
induced by ${\bf\Omega}_e$. These transformation rules can then be used 
to determine the relation between $F_\Lambda$ and the 
coordinates $X^\Lambda$. 
In components, the symplectic transformations are given by 
$$\eqalign{X &\rightarrow {\bf U} X+{\bf V}{\bf\eta}\partial F,\cr 
\partial F&\rightarrow -\eta {\bf V} X+\eta {\bf U}{\bf\eta}\partial F,}
\eqn\tra$$
where $X$ and $\partial F$ are $(n+1)$-dimensional complex vectors 
with components $X^\Lambda$ and $F_\Lambda$ respectively. 
It is clear from the transformation relations \tra\ that $\partial F$
can be identified with  
$i\eta X,$ in which case, a holomorphic function $F$ exists 
and is given, in terms of the 
coordinates $X$, by
$$F={i\over2}X^t\eta X.\eqn\Newyork$$
With the above relation, the complex vector $X$ transforms as
$$X\rightarrow ({\bf U}+i{\bf V})X={\bf M}X.\eqn\dd$$
This transformation implies that the complex vector $X$ should be identified 
with the complex vector parametrizing the 
${\textstyle SU(1, n)\over\textstyle SU(n)\otimes U(1)}$ coset.
The set of complex coordinates parametrizing this coset 
satisfy the following relation 
$$\eqalign{&\phi^\dagger\eta\phi={1}, \qquad\hbox{where}\quad
\phi=\left(\matrix{\phi^0\cr\vdots \cr\phi^{n+1}}\right),}\eqn\nem$$
and are parametrized in terms of 
unconstrained coordinates $z^\alpha$ by [\lust]
$$\phi^0={1\over \sqrt Y}, \quad \phi^j={z^\alpha\over \sqrt Y}, \quad
\alpha=1, 
\cdots, n,\eqn\berlin$$
where $Y={1-\sum_\alpha z^\alpha\bar z^\alpha}.$
Here we identify $X$ with the complex vector ${1\over\sqrt2}\phi$. 
The special coordinates in this case are given by
$z^\alpha$ and thus $Z^0=1,$ $Z^\alpha=z^\alpha,$ and the K\"ahler potential is
given by
$$K=-log(1-\sum_\alpha z^\alpha{\bar z}^\alpha).\eqn\sophie$$

A different embedding of $SU(1,n)$ into $Sp(2n+2, {\hbox{\bf R}})$ lead in
general to
a different relation between $F_\Lambda$ and $X^\Lambda.$
In fact once an embedding $\Omega_e$ is specified,
then for symplectic transformations $S\in Sp(2n+2,{\hbox{\bf R}}),$ 
the matrix
$$\Omega'_e=S\Omega_e S^{-1},\eqn\ae$$ defines another  embedding
with a new section. As an example, consider the element $S_1\in
Sp(2n+2,{\hbox{\bf R}}),$
given by
$$S_1=\pmatrix{{\bf\Sigma}&{\bf 0}\cr{\bf 0}&{\bf\Sigma}},\eqn\sis$$
where $${\bf\Sigma}=
\pmatrix{{\bf\sigma}&{\bf 0}\cr{\bf 0}&{\bf 1}},
\qquad
\bf\sigma={1\over\sqrt2}\pmatrix{1&1\cr 1&-1}.\eqn\sunday$$ 
Using \ae\ and \sis\ we obtain another
embedding of $SU(1,n)$ into $Sp(2n+2{\hbox{\bf R}})$ given by          
$$\Omega_e'=\pmatrix{{\bf\Sigma}{\bf U}{\bf\Sigma}&{\bf\Sigma}
{\bf V}\eta{\bf\Sigma}\cr -{\bf\Sigma}\eta{\bf V}{\bf\Sigma}&
{\bf\Sigma}\eta{\bf U}\eta{\bf\Sigma}}.\eqn\nst$$
For this embedding, we define a new symplectic section $(X',\partial F')$
which
transforms as a vector under the action of $\Omega'_e.$ It is obvious that 
the two sections $(X,\partial F)$ and $(X',\partial F')$ are related by 
the following relations
$$X'={\bf\Sigma}X,\qquad {(\partial F)}'={\bf\Sigma}\partial F.
\eqn\skeleton$$
This in components lead to
$$\eqalign{X'^0&={1\over\sqrt2}(X^0+X^1),\cr
X'^1&={1\over\sqrt2}(X^0-X^1),\cr
X'^j&=X^j,\qquad j=2, \cdots, n,\cr
F'_0&={1\over\sqrt2}(F_0+F_1)={i\over\sqrt2}(X^0-X^1)=iX'^1,\cr 
F'_1&={1\over\sqrt2}(F_0-F_1)={i\over\sqrt2}(X^0+X^1)=iX'^0,\cr
F'_j&=F_j=-iX^j=-iX'^j.}\eqn\yael$$
{}From these relations, it can be easily seen that there exists a 
holomorphic function
$F'$ which can be expressed in terms of $X'$ by
$$F'={i}\Big(X'^0X'^1-{1\over2}\sum_{j=2}^n (X'^j)^2\Big).\eqn\nf$$
In this case 
$$Z'^0=1, \quad Z'^1={1-z^1\over 1+z^1},\quad Z'^j={{\sqrt2} z^j\over
1+z^1},\eqn\vivaldi$$
and the K\"ahler potential is given by
$$K=-log({Z'}^1+{\bar {Z'}}^1-\sum_i{{Z'}^j\bar{Z'}^j}).\eqn\sophie$$
Formally, given two sections connected by a symplectic transformation
$$\pmatrix{X'\cr\partial F'}={\bf\Omega}\pmatrix{X\cr\partial F}=
\pmatrix{{\bf A}&{\bf B}\cr {\bf C}&
{\bf D}}
\pmatrix{X\cr\partial F},\qquad {\bf\Omega}\in Sp(2n+2,{\hbox{\bf
R}}),\eqn\ausfrt$$
then it can be shown that [\trieste]
$$F'={1\over2}\pmatrix{X&\partial F}\pmatrix{{\bf C}^t{\bf A}
&{\bf C}^t{\bf B}\cr {\bf D}^t{\bf A}&{\bf D}^t{\bf B}}
\pmatrix{X\cr\partial F}.\eqn\france$$
Using \Newyork,\france\ and \sis, the holomorphic function 
$F'$ defined in \nf\ can be obtained.

Finally, we discuss the automorphic function of $CP(n-1,1)$, which in 
physical terms might serve as a non-perturbative superpotential for the
scalar fields, provided that the isometry group remains a nonperturbative 
symmetry. In [\eighteen] Ferrara $et\ al,$ based on symmetry arguments, 
proposed an expression for the automorphic superpotential $\Delta$ for 
a special K\"ahler manifold with a symplectic section $(X,\partial F)$, defined
by
$$\log\vert\vert\Delta\vert\vert^2
=-\Big[\sum_{M, N}\log{\vert MX+N\partial
F\vert^2}\Big]_{\hbox{reg}}.\eqn\np$$

The constant vectors $M, N$ define $Sp(2n+2,{\hbox{\bf Z}})$ orbits and  
transform in a conjugate
way to that of $(X, \partial F)$ in order for the expression \np\ to 
be invariant
under the symplectic transformations.  

Given the duality group $\Gamma$ of a model with $n$ physical scalar fields
parametrizing 
a special K\"ahler manifold, 
the vector $(X, \partial F)$ transforms
by
$$\pmatrix{X\cr \partial F}\rightarrow\Omega_\Gamma
\pmatrix{X\cr \partial F}=\pmatrix{{\bf A}'&{\bf B}'\cr {\bf C}'&
{\bf D}'}
\pmatrix{X\cr \partial F},\eqn\h$$
where the matrix $\Omega_\Gamma$ is the embedding of the duality $\Gamma$
group in $Sp(2n+2,{\hbox{\bf Z}}).$
For \np\ to be invariant under the duality transformations,
we get the following transformation for the numbers $(M, N)$
$$\pmatrix {M\cr N}\rightarrow
\pmatrix{{\bf D}'&-{\bf C}'\cr -{\bf B}'&{\bf A}'}
\pmatrix{M\cr N}.\eqn\conj$$
For the case at hand, the symplectic transformation are those obtained
by the embedding of  $SU(1,n)$ into $Sp(2n+2,{\hbox{\bf Z}}).$
Such an embedding can be given by $\Omega_e$ for which \conj\ gives the

following transformation
$$\pmatrix {M\cr N}\rightarrow
\pmatrix{\eta {\bf U}' \eta&\eta{\bf V}'\cr -{\bf V}'\eta& {\bf U}'}
\pmatrix{M\cr N},\eqn\ne$$
where ${\bf U}'$ and ${\bf V}'$ are integer valued matrices.

Using the expression of the prepotential corresponding to this embedding as
given in \Newyork, 
eq. \np\ gives the following expression for the automorphic 
superpotential 
$$\log\vert\vert\Delta\vert\vert^2
=-\Big[\sum_{M'}\log{\vert M'^tX\vert^2}\Big]
_{\hbox{reg}},\eqn\npr$$
where $M'$ is an (n+1)-dimensional complex vector with components given by
$$\eqalign{M'_0&=M_0+iN_0, \cr M'_i&=M_i-iN_i,\quad i=1,\cdots,
n.}\eqn\hamburg$$
Under the action of the symplectic transformations \ne, 
we have
$$X\rightarrow\Omega X,\qquad \Omega\in SU(1, n,{\hbox{\bf Z}}).\eqn\hand$$
It can be easily shown that the following transformations 
for the quantum numbers hold,
$$\pmatrix{M'_0\cr -M'_i}\rightarrow \Omega^*\pmatrix{M'_0\cr -M'_i}.\eqn
\poil$$
Eq. \hamburg\ thus gives an explicit relation between the numbers $M, N$ and
those parametrizing $SU(1,n, {\hbox{\bf Z}})$ orbits. 
An explicit form of the
automorphic function can be obtained by using a regularization procedure
such as the $\zeta$-function regularization scheme [\eighteen,\fre].

To conclude, we have analysed the cosets $CP(n-1,1)$ with regard to the
construction of the holomorphic function describing their special geometry.
The knowledge of this function enables one to construct the lagrangian
of the $N=2$ supergravity where the scalar fields of the vector
supermultiplets parametrize the coset $CP(n-1,1)$.
It should be mentioned, however, that our method always provides a relation
between $F_\Lambda$ and the coordinates $X^\Lambda.$ It may occur that 
for a particular embedding an 
$F$ function does not exist. As a matter of fact, in some physically
interesting cases, one needs a formulation of the theory in which an $F$
function does not 
exist. This is the case in the study of perturbative corrections of 
vector coupling 
in $N=2$ heterotic string vacua [\van,\wit], and in the study of supersymmetry
breaking of $N=2$ supersymmetry down to $N=1$ [\br].
In this case, the physical quantities in the new basis,
are obtained by performing a symplectic transformation on the physical
quantities in a system of coordinates where an $F$ function exists. 

\centerline{\bf ACKNOWLEDGEMENT}
The work of W. A. Sabra is supported by PPARC, S. Thomas is supported by  
the Royal Society of Great Britain, and N. Vanegas is funded by 
Colciencias and U. de Antioquia. 
\vfill\eject
\refout
\end